\newcommand{\be}{\begin{equation}}
\newcommand{\ee}{\end{equation}}
\documentclass[twocolumn,showpacs,superscriptaddress,amsmath,amssymb]{revtex4}
\usepackage{graphicx}
\usepackage{dcolumn}
\usepackage{bm}

\begin{document}
\title{Do shape invariant solitons in highly nonlocal nematic liquid crystals really exist?\\}

\author{Milan S. Petrovi\'c}
\affiliation{Institute of Physics, P. O. Box 68, 11001 Belgrade,
Serbia} \affiliation{Texas A\&M University at Qatar, P. O. Box
23874, Doha, Qatar}

\author{Aleksandra I. Strini\'c}
\affiliation{Institute of Physics, P. O. Box 68, 11001 Belgrade,
Serbia} \affiliation{Texas A\&M University at Qatar, P. O. Box
23874, Doha, Qatar}

\author{Najdan B. Aleksi\'c}
\affiliation{Institute of Physics, P. O. Box 68, 11001 Belgrade,
Serbia} \affiliation{Texas A\&M University at Qatar, P. O. Box
23874, Doha, Qatar}

\author{Milivoj R. Beli\'c}
\affiliation{Texas A\&M University at Qatar, P. O. Box 23874,
Doha, Qatar}

\begin{abstract}\noindent
We question physical existence of shape invariant solitons in three dimensional nematic liquid
crystals. Using modified Petviashvili's method for finding eigenvalues and eigenfunctions,
we determine shape invariant solitons in a
realistic physical model that includes the highly nonlocal nature of the liquid crystal system.
We check the stability of such solutions by propagating them for long distances. We establish that any
noise added to the medium or to the fundamental solitons induces them to breathe, rendering them practically unobservable.
\end{abstract}

\maketitle

The fundamental spatial optical soliton is a beam that propagates in a
nonlinear (NL) medium without changing its transverse profile \cite{yuri}. Such
shape-invariant solutions are easily identified in (1+1)-dimensional [(1+1)D] NL systems
because the inverse scattering theory \cite{lamb}
guaranties their existence.
The situation is less clear in the {\em multidimensional} and {\em multicomponent}
systems. No credible inverse scattering theory is formulated in more than one dimension
and even when the localized solutions are found, no
credible procedure for guaranteeing their stability is established (however, there are
many {\em credible} linear stability analyses or stability criteria or numerical procedures, but not
for {\em rigorously} proven stability.)
In fact, wave instability and the
collapse of solutions are overriding concerns in multidimensional NL
systems \cite{sulem}.
Additional compounding difficulties arise in the multicomponent {\em vector}
models or in the scalar {\em nonlocal} models
in which the medium response is driven by the
optical field itself. Such are the models describing the generation of
solitary waves -- nematicons -- in nematic liquid crystals (NLCs).

Nonlocality is an important characteristic of many NL media. A
highly nonlocal situation arises in a nonlocal nonlinear (NN)
medium when the characteristic size of the response is much wider
than the size of the excitation itself \cite{conti3,conti1}.
In NLCs both experiments \cite{hen,cyril} and theoretical
calculations \cite{beec,pecc} demonstrated that the nonlinearity
is highly nonlocal.

For more complex nonlinearities, numerical techniques are
necessary to determine the soliton solutions. Soliton profile
calculations in NN media have been presented in a number of
papers, see for example \cite{rotschild,rots,ye}. The existence
and stability of 2D solitons in media with NN response
was discussed in \cite{krol}; even a high degree of nonlocality
did not guarantee the existence of stable high-order soliton
structures \cite{yak}. Orientational nonlinearity in NLCs is highly
nonlocal but the NL response is not perfectly quadratic,
implying that if one launches a Gaussian beam into the cell
it is only possible to observe {\em breathing}
solitons \cite{conti1,strin}.

In some publications soliton profiles were calculated using
semi-analytical models \cite{conti3,minz,ren,zhang}. For the more
general vectorial model, in which the order parameter in NLC is not
constant, steady {\it elliptical} soliton profiles are found numerically
in \cite{beeck}. To determine such profiles, the
authors demonstrated that it is necessary to include all
three components of the optical electric field.

However, what is puzzling is that even though everybody agrees that
shape-preserving solitons {\it do} exist in highly nonlocal NLCs, practically nobody
cared to present them explicitly.
Experimental accounts profusely mention steady nematicons, but careful inspection of
all published figures reveals self-focusing oscillations. True, experimental
results may be of not much help in this regard, because all experimental
setups feature a few mm long cells, which cannot capture slow (if any) convergence to
a steady profile. In this paper we find a family of fundamental solitons for
the same model and the {\em same} parameters; we check their stability in propagation and demonstrate that
any small change in the input shape, as well as in the medium, leads to the soliton breathing.
Consequently, we question the real physical observability of such shape-invariant solitons.

\begin{figure}\vspace{0mm}
\includegraphics[width=70mm]{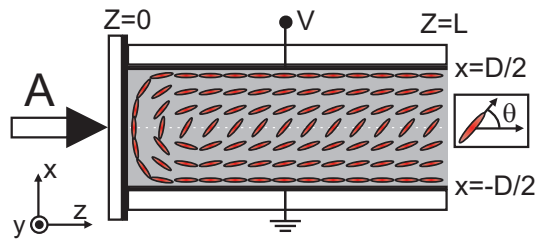}\vspace{0mm}
\caption{\label{Fig1} Liquid-crystal cell model adopted.}
\vspace{0mm}
\end{figure}

To find an exact fundamental soliton solution in a NLC model that
is \textit{not} vectorial, we use an iterative numerical
eigenvalue technique. We consider widely accepted scalar model of
beam propagation that is well suited for uniaxial NLCs and
low-intensity optical fields, which correspond to most situations
of practical interest. Also, we discuss the influence of boundary
conditions (BCs) on the shape and power of solutions, and analyze
soliton and Gaussian propagation using two different propagation
methods.

We adopt the well-known NN 3D scalar model in
NLCs, which provides good agreement with experimental data \cite{cyril}.
The optical beam polarized along the $x$ axis propagates in the
$z$ direction, while the NLC molecules can rotate in the $x$-$z$
plane. The liquid-crystal cell of interest is
sketched in Fig. \ref{Fig1}. The total orientation
of molecules with respect to the $z$ axis is denoted as
$\theta(x,y,z)$, whereas the orientation induced by the static
electric field only is denoted by $\theta_0$ (the pre-tilt angle).
The bias field points in the $x$ direction and is uniform in the
$z$ direction; hence the pre-tilt angle is uniform along the $z$
axis as well. The quantity $\hat{\theta}=\theta-\theta_0$
corresponds to the optically induced molecular reorientation.

The system of equations of interest consists of the scaled NL
Schr\"{o}dinger-like equation for the propagation of the optical
field $A$, and the diffusion equation for the molecular
orientation angle $\theta$ \cite{cyril,beec,strin}:

\begin{eqnarray}
\label{modelA} 2i\frac{\partial A}{\partial
z}+\triangle_{x,y}A+\alpha[\sin^{2}\theta-\sin^{2}\theta_{0}]A &=&0
, \\
\label{modelB}
2\triangle_{x,y}\theta+[\beta+\alpha|A|^{2}]\sin(2\theta)&=&0 ,
\end{eqnarray}
\noindent where the coefficients $\alpha$ and
$\beta$ are proportional to the optical and static permittivity
anisotropies of the NLC molecules, respectively.
Hard boundary conditions
(BCs) on the molecular orientation at the NLC cell faces
in the $x$ direction are assumed:
${\theta(x=-D/2,y)=\theta(x=D/2,y)}={\rm const.}$ \cite{marc}.
In our calculations we use data
corresponding to typical experimental conditions \cite{beec,strin,marc}.

\begin{figure}\vspace{0mm}
\includegraphics[width=\columnwidth]{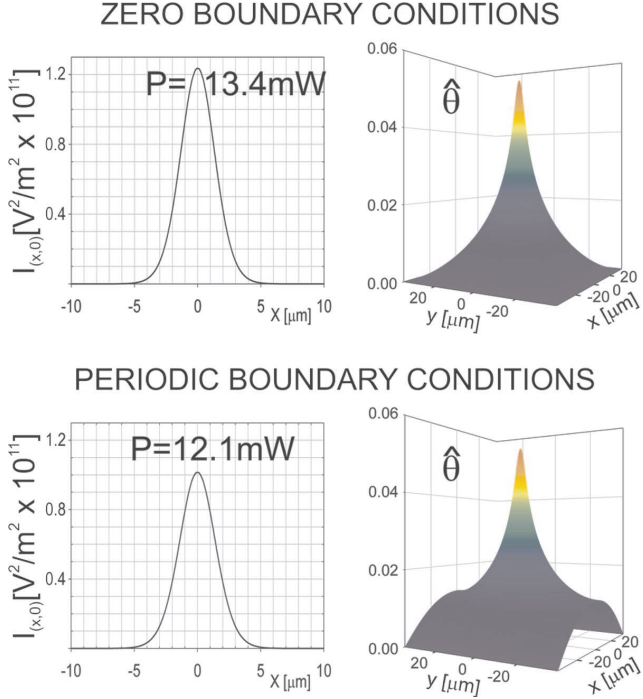}\vspace{0mm}
\caption{\label{Fig2} Fundamental soliton solutions for $\mu = 5
L_D^{-1}$. Intensity profiles and reorientation angle
distributions are shown for two different BCs. Other parameters
are the same.} \vspace{0mm}
\end{figure}

The solitary eigen-solutions are determined from the system of
Eqs. (\ref{modelA},\ref{modelB}) using the
modified Petviashvili's iteration method \cite{petv,yang,jovic}.
Equation (\ref{modelA}) suggests the existence of a fundamental soliton
of the form $A=a(x,y)e^{i \mu z}$, where $\mu$ is
the propagation constant. The real-valued function $a(x,y)$
satisfies the equation: $-\triangle a + (2 \mu + \mathbb{P}) a = \mathbb{Q}$,
where $\mathbb{P}=\alpha \sin^{2}(\theta_{0})$ and
$\mathbb{Q}=\alpha \sin^{2}(\theta) a$. After Fourier
transforming the equation for $a$, we get:

\be \label{iter0} {
\overline{a} = \frac{1}{|\mathbf{k}|^2 + 2 \mu}
(\overline{\mathbb{Q}} - \overline{\mathbb{P}a})} , \ee
where overbar denotes Fourier transform.
Straightforward iteration of Eq. (\ref{iter0}) does not converge
in general, so the stabilizing factors
had to be introduced \cite{yang,jovic}.
In each iteration step of Eq. (\ref{iter0}),
Eq. (\ref{modelB}) is treated using
a successive overrelaxation (SOR) method, until convergence is achieved.

\begin{figure}\vspace{0mm}
\includegraphics[width=70mm]{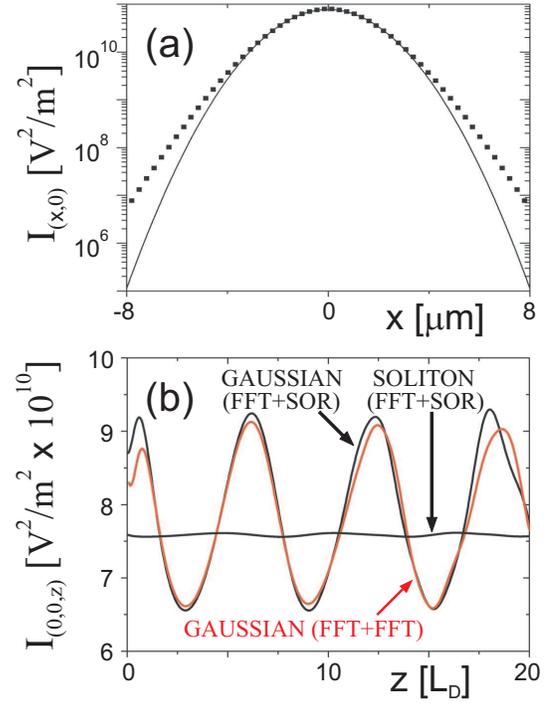}\vspace{0mm}
\caption{\label{Fig3} (a) Fundamental soliton intensity profile
obtained by the eigenvalue method (black dots), fitted with a
Gaussian. Parameters: $P$ = 10.6 mW, $\mu = 3.84 L_D^{-1}$; zero
BCs. (b) Soliton and Gaussian propagation using two different
propagation methods, FFT+SOR and FFT+FFT. The soliton power $P$ =
10.6 mW; Gaussian power $P$=10.6 mW for FFT+SOR, $P$=10.1 mW for
FFT+FFT.} \vspace{0mm}
\end{figure}

Stable soliton solutions are presented in Fig. \ref{Fig2} for two
different BCs. The shape and the power of fundamental
shape-invariant solutions naturally depend on the BCs applied.
Zero BCs ($\hat{\theta}=0$ on all boundaries) correspond to the
Dirichlet BCs. Periodic BCs correspond to the mixed BCs --
Dirichlet along the $y$ axis and Neumann along the $x$ axis. The
solution with the periodic BCs is more appropriate to the geometry
of the problem; furthermore, it is more acceptable on physical
grounds. The fundamental soliton so obtained requires less beam
power for the same value of the propagation constant and identical
other parameters. In identical conditions (but for BCs) the
solution requiring less power should be favored. The often used
$\theta_0 = \pi/4$ approximation leads to the solution with zero
BCs, and consequently such a soliton is less appropriate.

Spatial solitons in highly nonlocal media with quadratic response
possess Gaussian profiles \cite{conti3,conti1}.
However, the fundamental soliton profile is not Gaussian. The soliton intensity profile
compared to a Gaussian is shown in Fig. \ref{Fig3}(a); the difference is confined to the tails.
To check the stability of fundamental solitons,
we propagate them numerically; the results are presented in Fig. \ref{Fig3}(b).
Also included in Fig. \ref{Fig3}(b) is a case presenting propagation of a Gaussian
with similar parameters, but obtained using two different numerical methods.
In both methods a split-step beam propagation procedure based
on the fast Fourier transform (FFT) is used for the propagation of the
optical field. In the first method the diffusion equation for the
optically induced molecular reorientation is treated using the SOR method.
In the second method the diffusion equation is treated using the split-step
procedure again. One can see that the methods provide similar results; however the first method
is more accurate.

The problem with the FFT procedure is that it treats an array of {\it transversely}
periodic cells. Since the molecular reorientation is wide, it tends to slightly spill over
into the adjacent cells, i.e. back onto itself, adding to the optical field.
This is not an overriding problem in the propagation of a Gaussian,
as it only leads to a slightly amplified oscillation of the breathing solution.
However, it makes huge difference in the propagation of the fundamental soliton -- it
makes it impossible for the field to keep the shape-invariant input profile and therefore should be discarded.
Even the SOR solution slightly oscillates at lower accuracy; this, however, becomes imperceptible as the accuracy
is improved. In Fig. \ref{Fig3}(b) we show a case where the oscillation of the amplitude is still perceptible.
This brings us to an important point.

\begin{figure}\vspace{0mm}
\includegraphics[width=70mm]{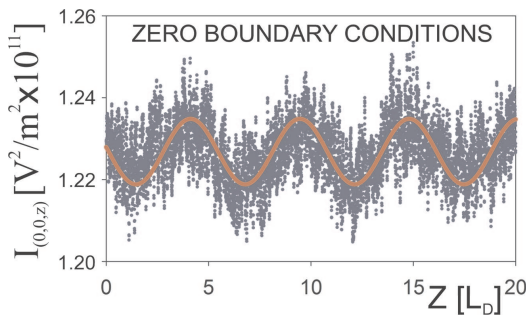}\vspace{0mm}
\caption{\label{Fig4} Propagation of the fundamental soliton in a
noisy medium. An amount of 0.5\% randomly changing noise is added
to $\theta_0$. Red sinusoidal fit is to guide eye.} \vspace{0mm}
\end{figure}

When one considers the propagation of a Gaussian beam using the two
propagation methods, the results are close. The propagation of Gaussians invariably leads to
breathing beams, regardless of the method of integration.
By the same token, when the fundamental soliton is propagated through the medium
in which a small random noise added to the underlying molecular orientation $\theta_0$,
a breathing solution is also obtained. The same phenomenon
happens as well when a small intensity noise is added to the fundamental profile, but $\theta_0$ kept unchanged.
This phenomenon is confirmed in our computations (Fig. \ref{Fig4}) and is not difficult to understand. In a highly NN medium any additional
energy from noise, no matter how small, cannot be radiated away and the solution has no way to relax to
the fundamental soliton. Therefore, it keeps oscillating about the fundamental
soliton, forming a {\it stable} breathing soliton. Since noise is unavoidable
in any realistic set-up, be it experimental or numerical, this fact
opens the question of the physical
observability of shape-invariant fundamental solitons in highly NN media.

In conclusion, we have presented calculations of shape-invariant
fundamental solitons in a highly nonlocal 3D scalar NCLs, for a
realistic physical model. Using modified Petviashvili's iterative
scheme we numerically determined the fundamental spatial soliton
profiles and found a family of solutions, depending on BCs.
We depicted stable propagation of such solitons.
We then demonstrated that upon propagation,
any amount of noise transforms the fundamental soliton into a breather,
making shape-invariant nematicons practically unobservable.

\vspace{-4mm}
\begin{acknowledgments}
This work has been supported by the Ministry of Science of the
Republic of Serbia, under the projects OI 171033 and 171006, and
by the Qatar National Research Foundation NPRP projects 25-6-7-2
and 09-462-1-074. Authors gladly acknowledge supercomputer time
provided by the ITS Research Computing group of the Texas A\&M
University at Qatar.
\end{acknowledgments}{}


\vspace{0pt}



\begin{thebibliography}{}

\bibitem{yuri} Yu. S. Kivshar and G. Agrawal, \textit{Optical solitons: From fibers to
photonic crystals} (Academic, San Diego, 2003).

\bibitem{lamb} G.L. Lamb, {\em Elements of Soliton theory} (John Wiley and Sons, New York, 1980).

\bibitem{sulem} C. Sulem and P. Sulem, {\sl The nonlinear
Schr\"odinger equation: Self-focusing and wave collapse} (Springer, Berlin, 2000).


\bibitem{conti3} C. Conti, M. Peccianti, and G. Assanto, Phys. Rev. Lett.
\textbf{91} (2003) 073901.

\bibitem{conti1} C. Conti, M. Peccianti, and G. Assanto, Phys. Rev. Lett.
\textbf{92} (2004) 113902.

\bibitem{hen} J.F. Henninot et al., J. Opt.
A \textbf{9} (2007) 20.

\bibitem{cyril} X. Hutsebaut et al.,
J. Opt. Soc. Am. B \textbf{22} (2005) 1424.

\bibitem{beec} J. Beeckman et al., Opt. Express \textbf{12} (2004)
1011.

\bibitem{pecc} M. Peccianti, C. Conti, and G. Assanto, Opt. Lett.
\textbf{30} (2005) 415.

\bibitem{rotschild} C. Rotschild et al., Opt. Lett.
\textbf{31} (2006) 3312.

\bibitem{rots} C. Rotschild et al., Phys. Rev. Lett.
\textbf{95} (2005) 213904.

\bibitem{ye} F. Ye et al., Opt. Lett.
\textbf{34} (2009) 2658.

\bibitem{krol} S. Skupin et al.,
Phys. Rev. E \textbf{73} (2006) 066603.

\bibitem{yak} A. I. Yakimenko, Y. A. Zaliznyak, and Yu. S. Kivshar,
Phys. Rev. E \textbf{71} (2005) 065603(R).

\bibitem{strin} A. Strini\'c et al., Opt.
Express \textbf{17} (2009) 11698.

\bibitem{minz} A. Minzoni, N. Smyth, and A. Worthy,
J. Opt. Soc. Am. B \textbf{24} (2007) 1549.

\bibitem{ren} H. Ren et al., J. Opt.
A \textbf{10} (2008) 025102.

\bibitem{zhang} H. Zhang, D. Xu, and L. Li, J. Opt.
A \textbf{11} (2009) 125203.

\bibitem{beeck} J. Beeckman et al., Opt.
Express \textbf{18} (2010) 3311.

\bibitem{marc} X. Hutsebaut et al.,
Opt. Commun. \textbf{233} (2004) 211.

\bibitem{petv} V. I. Petviashvili, Plasma Phys. \textbf{2} (1976) 469.

\bibitem{yang} J. Yang et al.,
Stud. Appl. Math. \textbf{113} (2004) 389.

\bibitem{jovic} D. Jovi\'c et al., Opt. Lett.
\textbf{32} (2007) 1857.

\end{thebibliography}
\end{document}